%% file: main.tex
\documentclass[10pt,twocolumn]{article}
\usepackage[letterpaper,margin=1in]{geometry}
\usepackage{amsmath,booktabs,graphicx,microtype}
\usepackage[T1]{fontenc}
\usepackage[hidelinks]{hyperref}
\hypersetup{pdftitle={Which Archived Results Survive a Rule Change? A Case Study in Revalidating Decisions},pdfauthor={Zengjian Liu, Tommy He, ELE Zhao, HongLiang Sun}}
\title{Which Archived Results Survive a Rule Change?\\A Case Study in Revalidating Decisions}
\author{%
  Zengjian Liu\\Peachy Lab \and
  Tommy He\\Peachy Lab \and
  ELE Zhao\\Peachy Lab \and
  HongLiang Sun\\Peachy Lab
}
\date{}
\begin{document}
\maketitle

\begin{abstract}
A research workflow can retain a correct numerical result after the evidence or approval that supported its use has expired. Common incremental build methods identify reusable computations, while deciding what remains usable also requires knowing which policy justified each approval. We examined a recorded parser migration across 312 paired research records, replayed both parsers with the same inputs, applied a calculation-file omission to the complete cohort, and evaluated current proof in a GNU Make comparison. Treating the prior projections as current would change at least one field in 11 of 312 records, and controlled cases show that even matching values can lack a current proof. The study separates numerical recovery from applicable approval and sets out the conditions under which a maintained dependency graph can preserve both.
\end{abstract}

\section{Introduction}

A research team may keep a numerical bound after replacing the rule used to select it. The team still has to decide whether the bound can be used in a current decision. For example, a parser can recover a bound from its saved calculations, while a prior approval covers a different input path or policy. The value, the record that justifies it, and the approval needed for use can change separately. A migration that tests only for unequal values can miss that distinction.

We studied an archived migration in a research workflow that paired reports, reviews and calculation files to produce an action bound and an open flag. Two archived parsers read the same paired records under different versions of the interpretation. The newer implementation sometimes required a selected calculation file. Later acceptance records also distinguished successfully computed bounds from those actually adopted. These records motivate the question of what a maintained workflow can carry forward after a rule change.

Our contribution is to make the distinction between unchanged output and current grounds for using it checkable. The full-cohort replay identifies which fields can change and how saved calculations contribute. A deterministic four-stage reference and a GNU Make implementation isolate what an explicit approval dependency, a valid parent chain and the retention of old records contribute. The reuse procedure follows established verifying traces.

\section{Related work}

Build systems separate task ordering from the decision to rebuild. Verifying traces bind outputs to dependency digests \cite[Secs.~2.3, 4.2]{build}. Shake can discover dependencies during a build \cite{shake}. Pluto formalizes sound and minimal rebuilding relative to tracked dependencies \cite{pluto}. Sotiropoulos et al.\ identify mismatches between build declarations and observed file operations \cite{buildfs}. Rattle analyzes when optimized forward builds agree with sequential execution \cite{rattle}. NixOS uses immutable build outputs \cite{nixos}, an established way to keep snapshots distinct. Bazel's Skyframe can reuse dependents after an invalidated node recomputes to the same value \cite{skyframe}. Our reuse check likewise follows the values a consumer reads, with an additional check on its chain of evidence.

Adapton tracks dependencies for demanded computations \cite{adapton}. Nominal Adapton adds stable names to identify work to reuse across executions \cite{nominal}. Acar et al.\ give consistency and correctness conditions for change propagation with memoization \cite{acar}, and differential dataflow extends incremental updates to nested iterations \cite{differential}. Regression test selection also depends on identifying affected work. Rothermel and Harrold compare control-flow graphs to identify tests that may exercise changed behavior \cite{regression}. Ekstazi uses files read during tests \cite{ekstazi}. Graves et al.\ evaluate selection cost and fault detection separately \cite{graves}. We similarly count work and check its outcome separately, with current validity assessed for each resulting stage.

Data provenance distinguishes the inputs that explain a result from the source locations of its values \cite{whywhere}. PROV represents entities, activities, and agents, including revisions of an entity \cite{prov}. ReproZip captures runtime dependencies for replay \cite{reprozip}, and noWorkflow records how scripts produced their results \cite{noworkflow}. CWLProv packages workflow definitions and execution records \cite{cwlprov}. Workflow Run RO-Crate describes the inputs, outputs and steps of a run \cite{crate}. These forms of provenance provide ways to record the history that a later approval policy can inspect.

Reproducible builds can check correspondence between source and binary by matching independent build outputs \cite{reprobuild}. An in-toto layout lets consumers check required steps and their authorized functionaries \cite{intoto}. SLSA also checks recorded build parameters against consumer expectations \cite{slsa}. In our setting the consumer checks the selected policy and the evidence chain for an archived research decision.

\section{Study design}

\subsection{Archived migration}

We joined the two archived parser outputs at their common cutoff by stable record identity. The analysis included all 312 identities with both parser versions, regardless of whether an earlier comparison showed a difference. For each identity we copied the bound report and review, the recorded calculation files and the available batch-version file into a temporary directory. The two code versions and the pinned report, review and calculation bytes matched recorded digests. Omission executions used the same copied report, review and batch bytes, leaving out the calculation files. We also checked the earlier parser under omission as a control. All 312 records were specified before the full-cohort replay. Earlier archival work had seen a subset of the changes.

We compared three fields in both contrasts. The bound could have a different amount, become available or unavailable, retain a value, or be absent on both sides. We also compared the open flag and the parser's bound-source attribution. Carrying the older result forward was evaluated against the newer parser on complete inputs. The omission contrast compared the newer parser with and without the calculation files. The open flag is a projected field. For three records with linked companion acceptances we separately compared the computed, accepted and installed bounds.

\subsection{Current-use check}

\begin{figure}[t]
\centering
\includegraphics[width=\columnwidth]{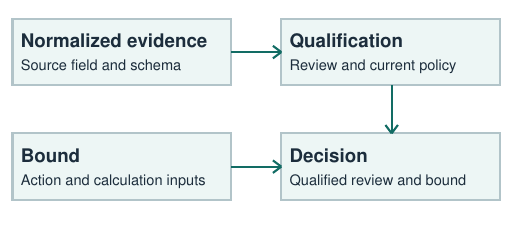}
\caption{Reference stages and direct workflow edges. The current policy is an explicit input of qualification. Each synthetic case supplies four stage outputs.}
\label{fig:workflow}
\end{figure}

The reference computes four outputs (Figure~\ref{fig:workflow}). A record contains its body, its body digest, the digests of projected inputs and the identity of the producing snapshot. Qualification depends on a review record, normalized evidence and the current approval policy. The decision requires a bound and qualification, and a supported result can be eligible, held or unresolved. For current use the record must match the declared inputs and have verifiable parents. An unresolved result can satisfy that contract when the policy leaves the read evidence insufficient.

With deterministic stage functions and complete declared inputs, a topological revalidation that either regenerates a stage from its current inputs or verifies the same inputs yields the full-rebuild value by induction \cite{build}. The proof additionally depends on the validity of the stored result and equivalence of the executable semantics.

In the Make comparison, four targets call the same stage functions as the content-based reference. We used GNU Make 3.81 with its standard prerequisite and target modification-time rules \cite{make}. The declared-policy arm names the policy file as a qualification prerequisite. A negative control omits that edge while the recipe still uses the policy. Every arm uses a separate historical copy with its digest checked. Make's working targets can be overwritten during a rebuild. A separate checker reconstructs the producing snapshot, payload digest and current input and parent bindings, so a retained value and a current proof receive separate scores. The supplementary comparison binds the exact approval bytes to a third-party time through an RFC~3161 token \cite{timestamp}.

\subsection{Use of AI}

The research owner chose the questions and study scope. AI tools assisted preparation of synthetic cases and protocols, implementation, analysis, verification, manuscript revision, and figure production. Synthetic cases and expected outcomes were fixed before execution in a separate session using the same model family as the work session. Earlier critiques of an anonymized brief used calls reported as MiniMax-M3 and grok-4.7-build and informed interpretation. Archived review records include model-assisted research decisions. Deterministic code executed and scored both Make comparisons without model or network requests.

\section{Results}

\subsection{When old parser outputs change}

\begin{figure*}[t]
\centering
\includegraphics[width=\textwidth]{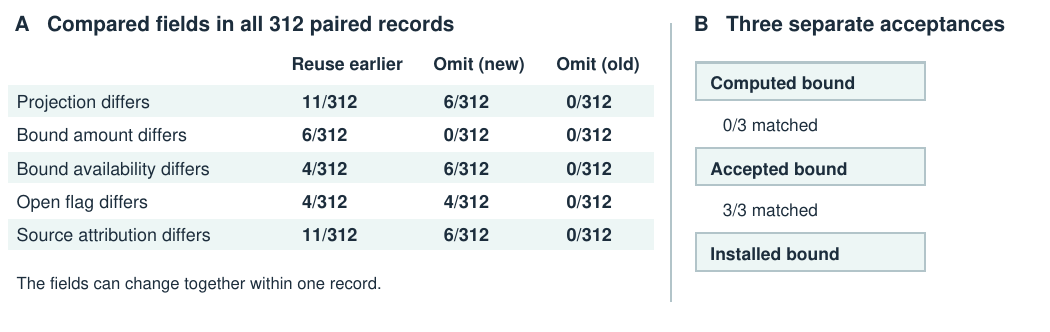}
\caption{\textbf{Recorded parser migration and separate acceptance.} A compares earlier versus newer outputs on complete inputs, and complete versus omitted calculation files within each parser. Each contrast covers all 312 paired archived records. Cells count identities with a change in the named field. B compares bound equality at each link for the three cases with a separate acceptance. The study archive retains the aggregate and unlinked category observations.}
\label{fig:historical}
\end{figure*}

Reusing the old projection would make a difference for 11 of 312 archived identities (Figure~\ref{fig:historical}A). Both parser versions matched all 312 projections attributed to them in the archive. Eleven records changed source attribution, six changed the numerical bound, and four changed whether a bound was available. Three of those four gained a bound and one lost it. Four also changed the open flag. One of the 11 had equal normalized amounts and an unchanged open flag while its attribution changed, showing that numeric equality can hide a change relevant to reproducibility.

The full-cohort omission removed a previously available bound in six of 312 cases. In each of those cases source attribution changed, and in four the open flag changed. The other 306 records had the same three-field output. The 301 identities outside the previously inspected subset all retained the same projection under omission. Omission left all 312 projections unchanged under the earlier parser, which supports the interpretation that the newer parser depended on the available calculation files in these six cases. Even the record that initially changed only attribution lost its bound under omission.

The disposition changes illustrate a separate authority boundary (Figure~\ref{fig:historical}B). Three computed limits required companion acceptances. None matched its accepted bound, and all three accepted bounds matched their corresponding installed bounds. The installation record reports zero new orders. Adoption followed a separate disposition in these cases.

\subsection{Value equality and current proof}

A correct retained payload can fail the current-use check. In one synthetic case, the selected evidence field changed while producing the same normalized amount. The four old values matched the expected current values, yet the evidence basis and two descendants lacked current proof. After regenerating the evidence, the matching downstream inputs allowed two records to be reused. In a second case all values matched but the qualification basis was missing. A third case kept the same bound and became unresolved because its old approval used the prior policy. These mechanisms were observed in earlier synthetic cases fixed before execution. Their per-stage counts are in Appendix~\ref{app:early}.

\input{make-results.tex}

\subsection{The cost of an avoided call}

A work count helps when the skipped evaluations cost more than the added verification. We inspected 41 model-review attempts from the recorded maintenance work. The 36 successful attempts took a median of 737~seconds (range 336--1,458). The tool's reported token total had a median of 187,587 (106,081--287,820). Five failed attempts took 126--900 seconds and lack a token total. The durations run from attempt start to finish and some calls can overlap. A related set of 35 companion review packages records provenance and review verdicts. Its model usage, review duration and human approval time remain unobserved.

\begin{figure}[t]
\centering
\includegraphics[width=\columnwidth]{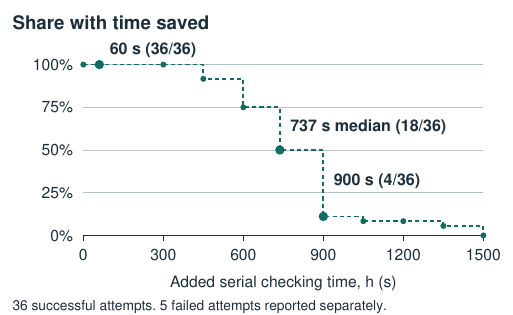}
\caption{\textbf{Added checks and conditional time savings.} Each dot gives the fraction of 36 successful review attempts whose elapsed time exceeded the added checking time. The step guide connects sampled thresholds. The scenario assumes one otherwise equivalent serial review is avoided. Five failed attempts are reported separately. Human waiting time was unobserved.}
\label{fig:review-costs}
\end{figure}

For a serial workload with fixed other work, the conditional saving is \(\sum_{j\in A}t_j-h\), where \(A\) consists of evaluations that are actually avoided and \(h\) is the added checking time. Figure~\ref{fig:review-costs} uses each successful historical attempt as one scenario for an equivalent review that could be avoided. With 60 added seconds, 36/36 scenarios save time (median 677 seconds saved). At the 737-second median, 18/36 save time. With 900 added seconds, 4/36 save time and the median is an added 163 seconds. Which real approval can be skipped and at what verification cost remain open. The deterministic kernel comparison uses a different work unit and appears in the appendix.

\section{Limitations and implications}

The archive is one selected historical migration. Two ancestor pointers agree between versions, while the corresponding files were unavailable at their recorded paths. The 300 rows with observed batch metadata resolve to one shared file. Its observed digest establishes the staged input used here. The earlier records do not pin that \mbox{digest.} The other 12 rows use the report-version fallback. The figures and unlinked category counts can be regenerated from the derived analysis files retained by the authors. Reproducing the historical parser executions requires the protected archive and the historical executable bytes.

The Make interventions deliberately exercise specific timestamp and integrity behavior, with the same declared local projections and evaluation functions. Their case counts describe those interventions. The study compares default timestamp checking with content and provenance checks in this implementation. GNU Make can also invoke content checks, represent policy inputs and use a separate history store. The recorded usage data cannot establish an end-to-end migration saving. An applicable approval and source of authority remain necessary for operational use.

\section{Artifact availability}

The authors retain a reproduction archive containing synthetic inputs, expected payloads and traces, stage code, the Make study, and the aggregate and category data underlying the historical and review-cost figures. The derived category blocks are sorted separately and contain no identity key. Historical source material, review texts, record identities, original bound amounts, and audit indexes remain in the restricted research archive.

\appendix
\input{supplement.tex}
\end{document}

%% file: make-results.tex
\paragraph{Make comparison.}
An explicit policy dependency was sufficient in the ordinary timestamp conditions. The 24 synthetic scenarios and expected stage values were fixed locally in a separate session before execution. The third-party timestamp on the approval was issued after all three repetitions. The six conditions each have four scenarios. The fresh-directory repetitions produced byte-identical traces, with all 288 planned case-by-strategy observations completed. We report the 24 scenarios as the study units.

\begin{table*}[t]
\centering\small
\caption{\textbf{Current results in the GNU Make comparison.}
Each ratio counts synthetic scenarios with all four stage values correct and their current proof chains valid. The omitted policy prerequisite is a negative control. Proof-only failures have four correct payloads and at least one current proof gap. Stage evaluations count one traversal of the 24 scenarios.}
\label{tab:make}
\input{generated-make-table.tex}
\end{table*}

The declared-policy Make arm delivered four correct, proved results in 18/24 scenarios (Table~\ref{tab:make}). Two cases returned an incorrect decision and four matched all values while failing current proof. When timestamps were retained on a changed policy, Make regenerated qualification in the one case whose review also acquired a newer timestamp. The other three in that condition kept an old decision or proof. When the old qualification basis was missing, a separately updated review likewise triggered the needed evaluation in one of four cases. A changed bound alone could trigger the decision without repairing its qualification ancestor. The separate proof check saw that gap even when the values matched.

Removing the policy prerequisite reduced current success to 14/24 and caused failure on all four ordinary policy edits. This shows why modelling the dependency matters even with valid timestamps. Full and content-based revalidation passed 24/24. Content checking reran only the evidence stage in each equal-evidence case, for four calls versus 12 under declared-policy Make. Across all conditions it ran 45 stages, compared with 47 for declared-policy Make and 96 for full rebuild. In qualification alone it ran 17 calls versus Make's 15, reflecting the work needed to repair some old proofs. Each strategy kept the same separate historical copy.

\paragraph{Boundary validation.}
Eight additional scenarios tested a change in the derivation of an equal bound, or retained timestamps on evidence or review inputs. A third-party timestamp bound the separate approval before execution (Appendix~\ref{app:time}). The three repetitions completed all 96 case-by-strategy observations with identical traces. Both Make arms passed 5/8, with two wrong decisions and one case of equal payloads but invalid proof. Both full and content-based revalidation passed 8/8. Policy was constant throughout these eight scenarios. Among four equal-bound scenarios, Make reran the bound and decision, eight calls in total, while content checking reran the bound alone, four calls. Of the four retained-timestamp scenarios, Make passed the one with an additional newer prerequisite. Across these eight scenarios content checking evaluated 13 stages, more than Make's 11, while obtaining the valid results.

%% file: generated-make-table.tex
\begin{tabular}{@{}lrrrr@{}}
\toprule
Condition & Full rebuild & Make with policy & Make without policy & Content checks \\
\midrule
Ordinary edit & 4/4 & 4/4 & 4/4 & 4/4 \\
Equal evidence value & 4/4 & 4/4 & 4/4 & 4/4 \\
Policy edit, newer timestamp & 4/4 & 4/4 & 0/4 & 4/4 \\
Policy edit, same timestamp & 4/4 & 1/4 & 1/4 & 4/4 \\
Missing old qualification basis & 4/4 & 1/4 & 1/4 & 4/4 \\
Numeric input change & 4/4 & 4/4 & 4/4 & 4/4 \\
\midrule
All cases & 24/24 & 18/24 & 14/24 & 24/24 \\
Proof-only failures & 0 & 4 & 5 & 0 \\
Wrong decisions & 0 & 2 & 5 & 0 \\
Stage evaluations & 96 & 47 & 39 & 45 \\
\bottomrule
\end{tabular}

%% file: supplement.tex
\section{Historical inputs and artifact}
\label{app:origin}

The candidate digests match their records and agree on the same 312 identities and cutoff. We verified the two parser files from local source objects and staged the same 624 report and review references and 940 calculation references for both. The omission used those copied report and review files and the available batch-version bytes. Two matching ancestor pointers lacked readable files at their original locations.

The derived category data retained by the authors contain 312 observations per measure and comparison. The values are sorted by category within each block and have no joining key. They reproduce the marginal comparisons, including the distinction between changes to an amount and changes to availability. The full joint observations, parser and source files require authorized local access. The three linked acceptances are reported separately.

\section{Earlier synthetic comparisons}
\label{app:early}

The earlier batches had five and 12 cases. The conceptual set, interface-compatible replacement and supplement were committed 0, 7\,min 34\,s and 16\,min 20\,s apart from the first commitment on one day. The plans were approved 18\,min 24\,s and 25\,min 41\,s after that first commitment. A late-source case used an adapter to select admissible input before evaluation. The computing arms each produced 68 correct, proved outputs, with 68, 64 and 37 evaluations under full, package and stage-level rebuild. Historical reuse would misstate 27/68 payloads if consumed as current, and its old proof failed on 40/68 outputs. An interrupted development run preserved five complete cases of 15 and a partial sixth.

\section{Timing of the protocols}
\label{app:time}

The 24-scenario protocol and labels were fixed locally before execution. The external timestamp on its approval was obtained after the runs. A separate eight-scenario protocol received a verified RFC~3161 timestamp before execution \cite{timestamp}. The first start was 90~seconds after the signed timestamp. The timestamp authority received a message imprint. The two studies use separate cases, labels and traces.

\section{Development timing}

Each of 24 arm orders revisited the same 15 development cases 25 times per arm (375 calls per arm and order). The in-memory interval covered stage evaluation and checking. It excluded file staging, post-run proof verification, scoring and obtaining reviews. One traversal used 60, 56 and 25 stage calls for full, package and content-based checks. The latter considered 60 current projections and visited 60 old nodes, with short-circuiting.

Content checks were slower in every pair. Their median excess was 16.20 microseconds per migration over full rebuild (14.76--17.28), and 7.00 over package rebuild (5.76--8.01). With stable verification and execution schedules, break-even increments would be 6.94 and 3.39 microseconds per avoided stage, respectively. These timings describe the deterministic kernel.